\documentclass[aps,prb,10pt,twocolumn,showpacs,floatfix,a4paper]{revtex4}
\usepackage{graphicx}
\usepackage[intlimits]{amsmath}
\usepackage{amssymb}

\begin{document}

\title{Density of states of a dissipative
quantum dot coupled to a quantum wire}

\author{Moshe Goldstein}
\author{Richard Berkovits}
\affiliation{The Minerva Center, Department of Physics, Bar-Ilan
University, Ramat-Gan 52900, Israel}

\begin{abstract}
We examine the local density of states of an impurity
level or a quantum dot coupled
to a fractional quantum Hall edge,
or to the end of a single one-dimensional Luttinger-liquid lead.
Effects of an Ohmic dissipative bath are also taken into account.
Using both analytical and numerical techniques we show that,
in general, the density of states exhibits
power-law frequency dependence near the Fermi energy.
In a substantial region of the parameter space
it simply reflects the behavior of
the tunneling density of states
at the end of a Luttinger-liquid, and is insensitive either
to the value of the dot-lead interaction or to the strength of dissipation;
otherwise it depends on these couplings too.
This behavior should be contrasted with the thermodynamic
properties of the level, in particular, its occupancy,
which were previously shown to depend on
the various interactions in the system
only through the corresponding Fermi edge singularity exponent,
and thus cannot display any Luttinger-liquid specific power-law.
Hence, we can construct different models,
some with and some without interactions in the wire (but with equal Fermi
edge singularity exponents), which would have very different level
densities of states, although they all result in the same
level population vs.\ energy curves.
\end{abstract}


\maketitle

\section{Introduction} \label{sec:intro}
Understanding the behavior of low-dimensional electronic systems has
been one of the main challenges of experimental and theoretical
physics in the last years.
These systems are important not only
as the basic building blocks
of nanoelectronic devices, but also for the
intricate strongly-correlated phenomena they exhibit.
An important subclass is that of
metallic (gapless) one-dimensional systems, whose
low energy dynamics is governed not by Fermi liquid theory,
but instead by the Luttinger liquid (LL) paradigm \cite{bosonization}.
This description applies to a wide variety of
experimental realizations, including narrow
quantum wires in semiconducting heterostructures, metallic nanowires,
and carbon nanotubes.
Closely related are chiral LLs,
formed at the edges of fractional quantum
Hall effect (FQHE) systems \cite{chang03}, and helical LLs,
the edges of spin quantum Hall insulators \cite{koenig08}.
The effect of impurities on these systems is interesting
from both the applicative and fundamental points of view.
These impurities could also be intentionally introduced,
in the form of, e.g., quantum dots and anti-dots.
Hence, there is no wonder that such questions have
attracted much effort recently.
However, most of these studies were restricted to
investigation of transport phenomena
\cite{bosonization,chang03,koenig08,kane92,furusaki93,furusaki98,
auslaender00,postma01,nazarov03,polyakov03,komnik03,lerner08,goldstein10b},
while other effects received much less attention
\cite{furusaki02,lehur05,sade05,weiss06,weiss07,wachter07,bishara08,
goldstein09,elste10}. 

In this work we study probably the most basic example of such a system,
namely, a single level in the vicinity of a fractional quantum Hall edge,
or, equivalently \cite{fn:fabrizio95},
a level attached to the end of a single LL wire \cite{furusaki02}.
We will refer to the two components (in both systems)
as ``dot'' and ``lead'' respectively.
We include in our treatment the effects
of short range dot-lead interaction,
as well as the influence of an Ohmic dissipative bath
(e.g., electromagnetic fluctuations in gate electrodes)
\cite{kamenev_gefen,lehur05,spin_boson,spin_boson2}.
In a recent work \cite{goldstein09} 
we have studied the thermodynamic properties of
the model (e.g., the level population, entropy, and specific heat),
and found that they are \emph{universal}, in the sense that
they depend on the various interactions in the model
(intra-lead, dot-lead, and dot-bath) only through a single
parameter, the Fermi edge singularity exponent.
Thus, thermodynamics can neither be used to identify non-Fermi
liquid behavior, nor to extract LL parameters.
In this work we proceed to study, both analytically and numerically,
the level density of states (LDoS),
which may be probed by tunneling or absorption spectroscopies.
We find that the LDoS is sensitive to LL physics,
even though its integral (times the Fermi function)
gives the level occupancy, which is universal in the above sense.
As we show below, the LDoS features power-law behavior near
the Fermi energy. For not too strong interactions
the exponent in this power-law
is actually determined by LL physics alone, and is independent
of the level-lead and level-bath interactions.
This and many other results derived below cannot be achieved using
perturbative calculations \cite{wachter07}.

The rest of this paper is organized as follows:
In Sec.~\ref{sec:model_cg} we present our model, and apply to it
the Anderson-Yuval Coulomb-gas (CG) expansion
\cite{yuval_anderson,schotte71,si93,kamenev_gefen,wiegmann78,fabrizio95}.
We then proceed to analytic
treatment of the LDoS in Sec.~\ref{sec:analysis}, and to
numerical calculations in Sec.~\ref{sec:numerics}.
Finally, we summarize our findings in Sec.~\ref{sec:conclude}.

\section{Model and Coulomb-gas expansion} \label{sec:model_cg}
The system is described by the
Hamiltonian $H=H_{D}+H_{L}+H_{DL}+H_{B}+H_{DB}$.
The first term is the dot Hamiltonian
$H_{D}=\varepsilon_0 d^{\dagger}d$, with $d^{\dagger}$ and $d$
the level creation and annihilation operators, respectively,
and $\varepsilon_0$ the level energy.
The second term is the lead Hamiltonian.
It can be written in the form
\begin{equation}
 H_L = \frac{v}{4\pi} \int_{-\infty}^{\infty} [\partial_x \phi(x)]^2 dx,
\end{equation}
using \emph{chiral} bosonic field $\phi(x)$ obeying the commutation
relation $[\phi(x),\phi(y)] = i \pi \text{sgn}(x-y)$,
where $v$ is the velocity of excitations \cite{fn:fabrizio95}.
The level and the lead are coupled by:
\begin{equation}
\label{eqn:hdl}
H_{DL} = 
t_0 d^{\dagger} \psi(0) +
\text{H.c.} + 
U_0
\left( d^{\dagger}d - \textstyle\frac{1}{2} \right)
\frac{\sqrt{g}}{\pi} \partial_x \phi(0).
\end{equation}
The two terms in this equation describe, respectively, dot-lead hopping
(with $t_0$ the tunneling matrix element),
and local dot-lead interaction whose strength is $U_0$.
The electronic annihilation operator at the end of the lead can be written as
$\psi(0)=\chi e^{i\phi(0)/\sqrt{g}}/ \sqrt{2\pi a}$,
where $\chi$ is a Majorana operator, $a$ is a short distance cutoff
(e.g., the lattice spacing), and $g$ is the LL interaction
parameter ($g<1$ for repulsion, $g>1$ for attraction).
For a FQHE system with filling $\nu$, $g=\nu$ for electron tunneling
(i.e., a dot outside the FQHE bar).
Finally, the level is coupled to a bath of harmonic oscillators
\cite{kamenev_gefen,lehur05,spin_boson,spin_boson2}
(describing, e.g., electromagnetic fluctuations in control gates),
governed by $H_{B} = \sum_k \omega_k a_k^\dagger a_k $.
The dot-bath coupling can be written as
$H_{DB}=\left(d^\dagger d - \frac{1}{2}\right) \sum_k \lambda_k (a_k^\dagger + a_k)$.
We assume Ohmic dissipation, i.e., linear low-frequency behavior of the
bath spectral function:
$J_B (\omega) \equiv \sum_k \lambda_k^2 \delta(\omega - \omega_k) = K \omega$.

We examine this model employing the Anderson-Yuval CG expansion
\cite{yuval_anderson}. In this approach, any quantity
of interest is expanded to all orders in $t_0$.
This results in a series of correlation functions,
which need to be evaluated for vanishing
$t_0$.\cite{yuval_anderson,schotte71,si93,kamenev_gefen,wiegmann78,fabrizio95}
The level-lead interaction gives rise to a potential at the end
of the lead,
which alternates between $U_0/2$ and $-U_0/2$ whenever
an electron tunnels in or out of the level.
Similarly, the $k$th bath oscillators experience a
shift in its equilibrium position, proportional to $\lambda_k$.
We thus have a sequence of Fermi edge singularity
events \cite{noziers69}.
The solution of this latter problem enables the
calculation of all the terms in the series of correlation functions.

\begin{figure}
\includegraphics[width=9cm,height=!]{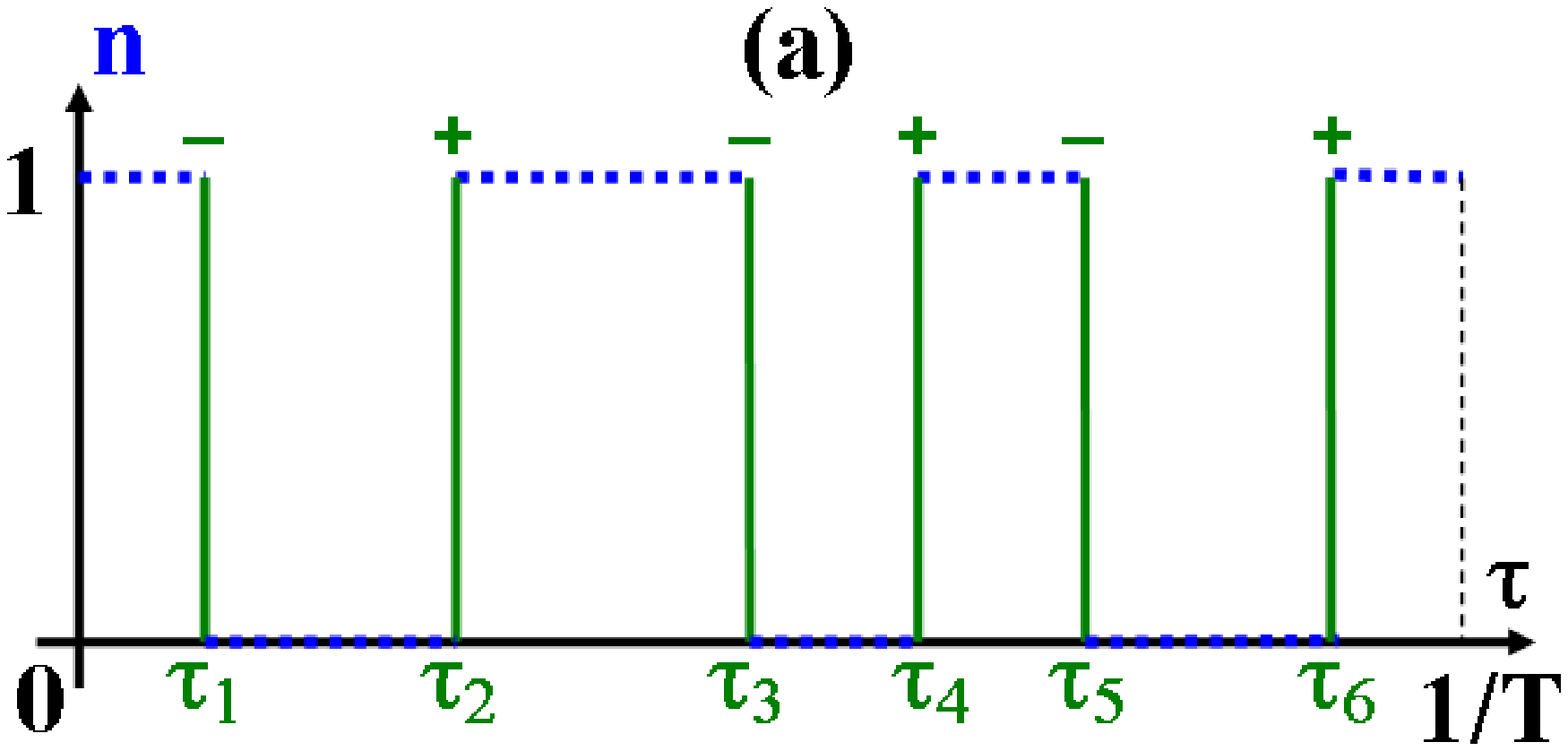}
\includegraphics[width=9cm,height=!]{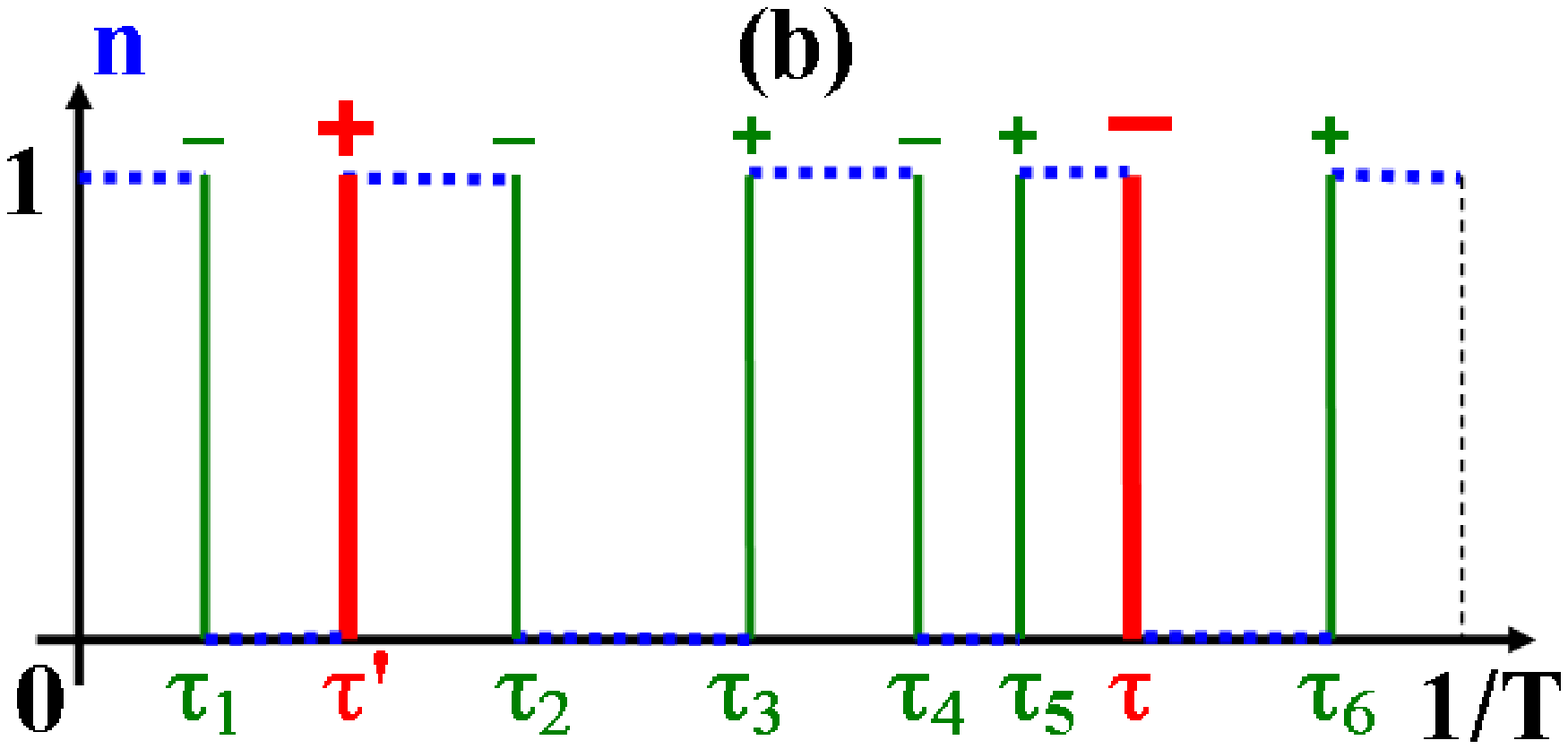}
\caption{\label{fig:cg}
(Color online)
A typical term (with $2N=6$ $t_0$-charges and $s=-1$)
in the CG expansions for
(a) the partition function [Eqs.~(\ref{eqn:cg_z1})--(\ref{eqn:cg_z2})];
(b) the dot Green function for $\tau>\tau^\prime$
[Eqs.~(\ref{eqn:cg_g1})--(\ref{eqn:cg_g2})]
(here $s^\prime=1$ and $M=0$ so a single $t_0$-charge precedes $\tau^\prime$,
whereas there are $2M^\prime = 4$ $t_0$-charges between $\tau^\prime$
and $\tau$).
Signs and positions of the charges are indicated,
with $t_0$-charges marked by thin green vertical lines,
and d-charges by wide red ones.
Level population as function of imaginary time
is denoted by horizontal dashed blue lines.
}
\end{figure}

Recently \cite{goldstein09} 
we have studied in this way the partition function $Z$ of the model,
whose derivatives with respect to the parameters of the system
(for example, the level energy $\varepsilon_0$ and the temperature $T$)
give us the thermodynamic properties
(e.g., the level population, entropy, and specific heat).
We were able to rewrite the series expansion for $Z$
in the form of a grand canonical partition function of a classical
system of particles. These represent hopping events
generated by $t_0$, and thus reside on
the imaginary time axis of the original quantum model,
which is a circle with circumference $1/T$.
Each particle is assigned a positive (negative) charge if
it represents tunneling of an electron from the lead to the dot
(from the dot to the lead).
Hence, there must be an even number of charges,
which have to appear in alternating order of signs.
The position of the $i$th particle is $\tau_i$, and
the sign of the charge of the first particle is denoted by $s$.
The partition function then reads:
\begin{multline} \label{eqn:cg_z1}
Z = \sum_{ \substack{N=0 \\ s=\pm1}}^{\infty}
y^{2N}
\int_{0}^{1/T} \frac{d\tau_{2N}}{\xi}
\int_{0}^{\tau_{2N}-\xi} \frac{d\tau_{2N-1}}{\xi}
\dots \\
\int_{0}^{\tau_3-\xi} \frac{d\tau_2}{\xi}
\int_0^{\tau_2-\xi} \frac{d\tau_1}{\xi}
e^{- S_{CG} (s, \{ \tau_i \})} ,
\end{multline}
The charges have a fugacity $y = \sqrt{\Gamma_0\xi/\pi}$,
where $\Gamma_0 = \pi |t_0|^2 \rho_L$ is the noninteracting level width
[$\rho_L = 1/(\pi v)$ is the corresponding lead local
density of states],
and $\xi \sim a/v$ is a short-time cutoff.
The CG action is given by:
\begin{multline} \label{eqn:cg_z2}
S_{CG}( s, \{ \tau_i \} ) = 
\sum_{i<j =1}^{2N} \vec{e}_i \cdot \vec{e}_j 
V_{C} (\tau_j-\tau_i)
+ \\
\varepsilon_0 \left[ \frac{1-s}{2T}
+ s \sum_{i=1}^{2N} (-1)^i \tau_i \right].
\end{multline}
The first term of this classical Hamiltonian describes an interaction between
the particles, with
$V_{C} (\tilde{\tau}) = \ln \{ \pi T \xi/ \sin [ \pi T |\tilde{\tau}| ] \}$.
This interaction is similar in form to 2D Coulomb interaction,
and is the origin of the name ``CG expansion''.
The charges are two component vectors, where the two components
correspond to the effects of the coupling with the lead and the bath,
respectively.
They are given by
$\vec{e}_i = s (-1)^{i-1} \vec{e}_0$,
where the squared-magnitude of the charges, to be denoted by
$\alpha_{\text{FES}} \equiv |\vec{e}_0|^2$, is the Fermi
edge singularity exponent of the model.
It is defined by behavior of the zero-temperature
correlator of $d^{\dagger} \psi(0)$ with its Hermitian conjugate,
calculated at $t_0=0$.
This correlator decays as $\tilde{\tau}^{-\alpha_{\text{FES}}}$
for long time $\tilde{\tau}$.
In our system we have found that
$\vec{e}_0 = \left( 1/\sqrt{g} - 2 \sqrt{g} \delta_\text{eff}/\pi, \sqrt{K} \right)$,
where $\delta_\text{eff}$ 
is the effective phase shift in the lead
due to the dot-lead coupling \cite{goldstein09}. 
It is equal to $U_0/(2 v)$ in straightforward bosonization,
but is more complicated in general.
It may be extracted from, e.g., finite-size energy differences,
which could be calculated either numerically or analytically
(via the Bethe ansatz) \cite{goldstein09}. 
The other part of the CG action accounts for the energetic
cost of $\varepsilon_0$ per unit imaginary time for
each interval in which the level is populated.
Its form is analogous to an electric
field applied to the classical system of charges.
A typical configuration is depicted in Fig.~\ref{fig:cg}(a).

A similar treatment can be given to the LDoS,
which we shall denote by $\rho_{D}(\omega)$.
It is equal to the imaginary part of the level retarded Green function
(multiplied by $-1/\pi$).
The retarded Green function is in turn
the result of analytic continuation of the Matsubara Green
function from the upper half of the complex frequency plane \cite{mahan}.
The latter Green function is defined by
$G_{D}(\tau-\tau^\prime) =
-\text{Tr}\{{\hat{T}}_{\tau} e^{-H/T} d(\tau) d^\dagger(\tau^\prime)\}/Z$,
where ${\hat{T}}_{\tau}$ is the imaginary time ordering operator. 
Following the same methods as above,
the numerator of this expression can also be given a CG
representation. This CG has the same form as
Eqs.~(\ref{eqn:cg_z1})--(\ref{eqn:cg_z2}), with two additional charges
of sizes $\pm \vec{e}_d$, $\vec{e}_d = \vec{e}_0 - (1/\sqrt{g},0)$,
inserted at $\tau^\prime$ and $\tau$, respectively.
These charges correspond to the level creation and annihilation
operators appearing in the definition of the Green function.
In the following we will refer to these as ``d-charges'',
to distinguish them form the other ``$t_0$-charges'',
which originate from the $t_0$ term.
The contribution of each such configuration is to be multiplied by
$\text{sgn}(\tau^\prime-\tau)$ to account for the Fermi statistics.
Thus, for $\tau>\tau^\prime$
the full CG expression for the dot Green function is:
\begin{widetext}
\begin{multline} \label{eqn:cg_g1}
G_D (\tau>\tau^\prime) = 
-\frac{1}{Z}
\sum_{ \substack{N=0 \\ s=\pm1}}^{\infty} y^{2N}
\sum_{ M = 0 }^{N - s^\prime}
\sum_{ M^\prime = 0 }^{N - s^\prime - M}
\int_{\tau+\xi}^{1/T} \frac{d\tau_{2N}}{\xi}
\dots 
\negthickspace \negthickspace \negthickspace \negthickspace
\int_{\tau+\xi}^{\tau_{2(M+M^\prime)+s^\prime+2}-\xi}
\frac{d\tau_{2(M+M^\prime)+s^\prime+1}}{\xi}
\times \\
\int_{\tau^\prime+\xi}^{\tau-\xi} \frac{d\tau_{2(M+M^\prime)+s^\prime}}{\xi}
\dots 
\int_{\tau^\prime+\xi}^{\tau_{2M+s^\prime+2}-\xi} \frac{d\tau_{2M+s^\prime+1}}{\xi}
\int_0^{\tau^\prime-\xi} \frac{d\tau_{2M+s^\prime}}{\xi}
\dots 
\int_0^{\tau_2-\xi} \frac{d\tau_1}{\xi}
e^{- S_{CG,D} (s, \tau, \tau^\prime, \{ \tau_i \})} ,
\end{multline}
where $s^\prime \equiv (1-s)/2$.
The first $2M+s^\prime$ $t_0$-charges
occupy the interval $[0,\tau^\prime]$,
the following $2M^\prime$ charges
reside in the interval $[\tau^\prime, \tau]$,
and the last $2(N-M-M^\prime) - s^\prime$ $t_0$-charges
are in the interval $[\tau, 1/T]$.
The classical action is given by:
\begin{multline} \label{eqn:cg_g2}
S_{CG,D}( s, \tau, \tau^\prime, \{ \tau_i \} ) =
|\vec{e}_0|^2 \sum_{i<j=1}^{2N} s_i s_j
V_C (\tau_j-\tau_i) + 
\vec{e}_0 \cdot \vec{e}_d \sum_{i=1}^{2N} s_i
\left[ V_C (\tau_i-\tau^\prime) - V_C (\tau_i-\tau) \right]
- |\vec{e}_d|^2 V_C (\tau-\tau^\prime) + \\
\varepsilon_0 \left[ \frac{1-s}{2T}
- \sum_{i=1}^{2N} s_i \tau_i + \tau - \tau^\prime \right],
\end{multline}
\end{widetext}
where the sign of the $i$th $t_0$-charge is
$s_i = s (-1)^{i-1} \text{sgn} (\tau_i-\tau^\prime) \text{sgn} (\tau_i-\tau)$.
A typical configuration is shown in Fig.~\ref{fig:cg}(b).
Similar expressions hold for $\tau<\tau^\prime$.

Comparing the two CG expansions, the following observation emerges:
the CG expansion for the partition function contains only three
parameters: $\Gamma_0$, $\varepsilon_0$, and $\alpha_{\text{FES}}$,
while expansion for the Green function depends on $g$ too
(through $\vec{e}_d$).
Hence, the different interaction types
(i.e., interactions in the wire, the dot-wire interaction,
and the dot-bath coupling) affect the partition function
through a single parameter,
the Fermi edge singularity exponent $\alpha_{\text{FES}}$.
Thus, thermodynamic measurements cannot be used to distinguish between
the different interaction types.
In other words,
one can construct very different models, whose interactions differ in
strength and even in sign, which will have the same thermodynamic properties,
provided $\Gamma_0$, $\varepsilon_0$, and $\alpha_{\text{FES}}$
are indeed the same \cite{goldstein09}. 
On the other hand, the LDoS, which depends \emph{explicitly}
on $g$, will exhibit different behavior for these different systems.
Hence, it can be used to extract the strength of intra-wire interactions,
as we show below.

\section{Analysis of the level density of states} \label{sec:analysis}

As noted in our earlier work \cite{goldstein09},
the CG obtained here is identical to the original
Anderson-Yuval expansion \cite{yuval_anderson}
for the anisotropic single-channel Kondo model \cite{hewson},
demonstrating that the models are equivalent \cite{furusaki02,lehur05}.
Under this mapping the level population becomes
the magnetization of the Kondo spin (plus one half).
Hence, $\varepsilon_0$ is analogous to a local magnetic field.
Similarly, $J_{xy}$ is related to $\Gamma_0$,
and $J_z$ to $\alpha_{\text{FES}}$.
The CG parameters obey the famous Kondo renormalization group (RG)
equations \cite{yuval_anderson,hewson}, which read, in our notations:
\begin{align}
 \label{eqn:rg1}
 \frac{d y}{d \ln \xi} & = \frac{2-\alpha_\text{FES}}{2} y, \\
 \label{eqn:rg2}
 \frac{d \alpha_\text{FES}}{d \ln \xi} & = - 4 y^2 \alpha_\text{FES}.
\end{align}
Thus, the system considered can be in one of two phases,
a strong coupling (antiferromagnetic-Kondo like) phase 
and a weak coupling (ferromagnetic-Kondo like) phase. 
The transition occurs, for small $\Gamma_0$, at
$\alpha_{\text{FES}}=2 + O (\sqrt{\Gamma_0 a/v})$.
In the weak-coupling phase the Coulomb charges form tightly-bound pairs.
The level is thus effectively decoupled at low energies,
resulting in its population being discontinuous as a function
of $\varepsilon_0$ at zero temperature \cite{furusaki02,lehur05,weiss07}.
In the strong-coupling phase free Coulomb charges proliferate.
The impurity is well-coupled with the lead,
so the level population is analytic in
$\varepsilon_0$, and could be extracted from
the Bethe ansatz solution of the Kondo problem \cite{hewson}.
In particular, for small values of $\varepsilon_0$ one has
$n(\varepsilon_0) - 1/2\sim - \varepsilon_0/T_K$,
where $T_K = (v/a) ( \Gamma_0 a/v )^{1/(2-\alpha_{\text{FES}})}$
is the Kondo temperature (effective level width, reducing
to $\Gamma_0$ in the noninteracting case) \cite{wachter07}.
Hence, in this phase the population does not exhibit any nontrivial
power-law dependence on $\varepsilon_0$ or $T$.
The same applies to other thermodynamic quantities.

What are the implications of this on the LDoS?
As we now show, we typically find that at zero temperature
we have a power-law behavior $\rho_D(\omega) \sim |\omega|^\delta$
in the vicinity of the Fermi energy, i.e., when $|\omega|$
is much smaller than $T_K$ in the strong-coupling phase,
and than the bandwidth $\sim v/a$ in the weak-coupling phase.
The values of $\delta$ in the different regimes are
summarized in Table~\ref{tbl:rho_d}.
It should be noted that when $T>0$,
or when the lead length $L$ is finite,
such power-law singularity will be smeared and become
$[\max(|\omega|, T, v/L)]^\delta$.
We will now consider each phase separately.

\begin{table}
\caption{ \label{tbl:rho_d}
Summary of the analysis of Sec.~\ref{sec:analysis}.
In the vicinity of the Fermi energy (when $|\omega|$ and $T$ are 
much smaller than $T_K$ in the strong-coupling phase,
and than the bandwidth $\sim v/a$ in the weak-coupling phase)
we have $\rho_D(\omega) \sim [\max(T,|\omega|)]^\delta$,
with the values of the exponent $\delta$ in the different regimes
denoted in the table below.
When two values are given, the smaller one
will give the dominant contribution \cite{fn:dominant_term}.
}
\begin{ruledtabular}
\begin{tabular}{ccc}
 Phase           & $|\omega| \ll |\varepsilon_0|$ & $|\omega| \gg |\varepsilon_0|$ \\ \hline
 Strong-coupling & $1/g-1$                    & $1/g-1$                    \\
 Weak-coupling   & $1/g-1$ 
 & $\alpha_\text{FES}^d-1$ or $1/g-1$    \\
\end{tabular}
\end{ruledtabular}
\end{table}

\subsection{The strong-coupling phase}

\begin{figure}
\includegraphics[width=9cm,height=!]{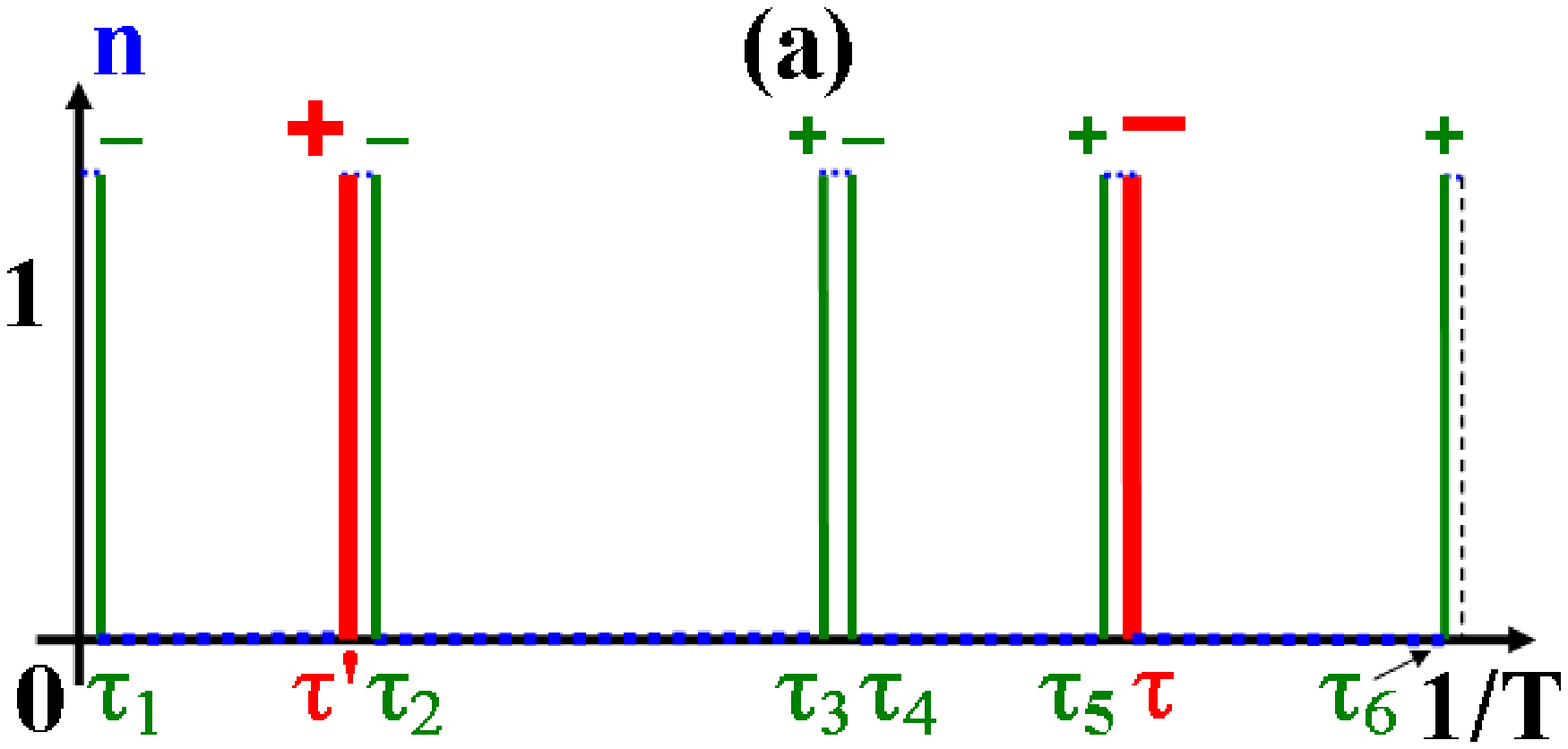}
\includegraphics[width=9cm,height=!]{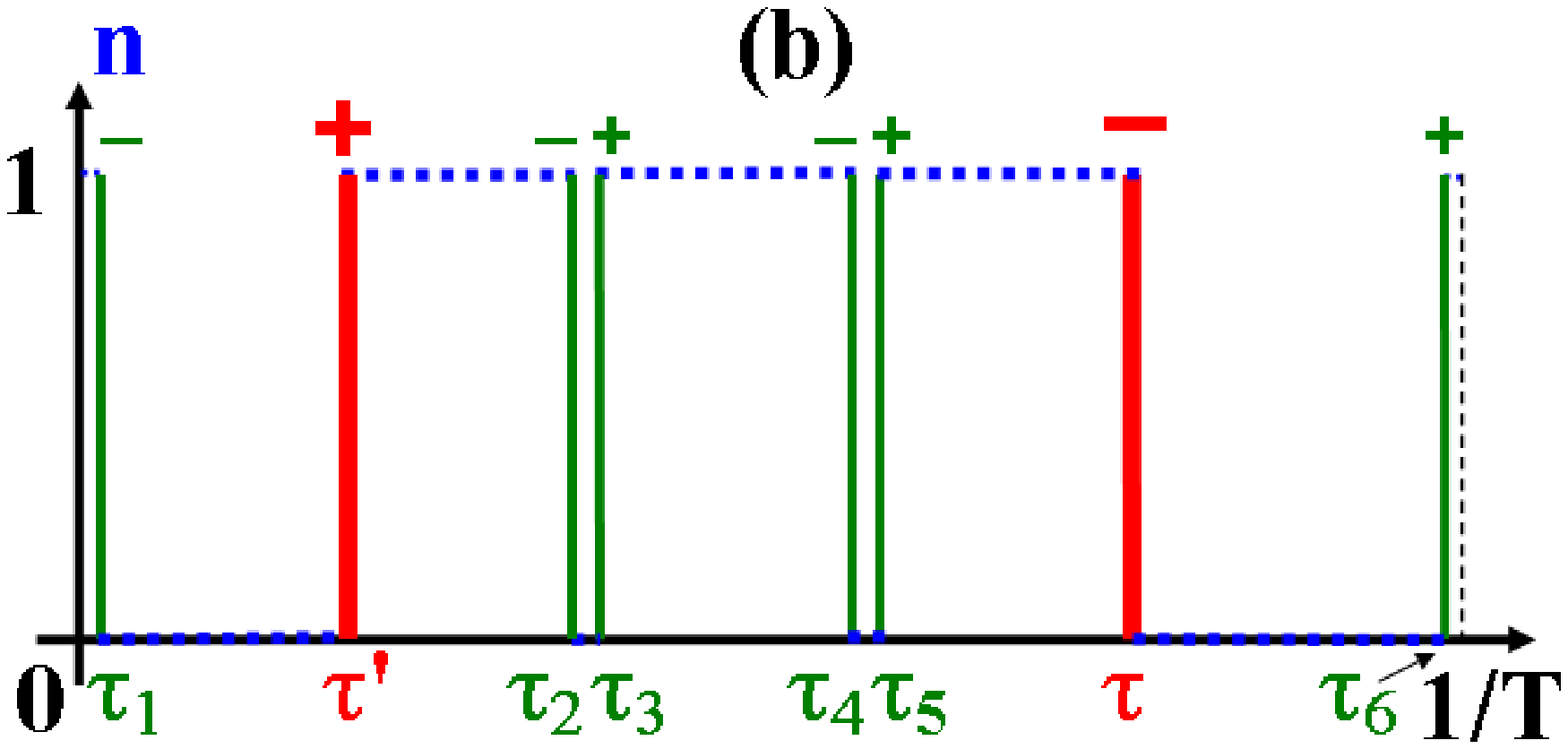}
\caption{\label{fig:cg_gp}
(Color online)
Typical configuration of the
CG expansion for the dot Green function [Eqs.~(\ref{eqn:cg_g1})--(\ref{eqn:cg_g2});
here $\tau>\tau^\prime$, $s=-1$, $2N=6$, $s^\prime=1$, $M=0$, and $2M^\prime = 4$,
as in Fig.~\ref{fig:cg}(b)] for
(a) large $|\varepsilon_0|$ in both phases
(and possibly small $|\varepsilon_0|$ in the weak-coupling phase \cite{fn:dominant_term}) ---
d-charges are screened; 
(b) small $|\varepsilon_0|$ in the weak-coupling phase ---
d-charges are not screened. 
Notations are the same as in Fig.~\ref{fig:cg}.
See the text for further details.
}
\end{figure}

Let us start from the strong-coupling phase. 
When $|\varepsilon_0|$ is large enough
(with respect to $T_K$),
CG charges must appear in tightly-bound pairs,
since large intra-pair separation
is suppressed by the level energy, as dictated by the last term of
Eq.~(\ref{eqn:cg_g2}).
The two d-charges added in the calculation of the level
Green function will also be accompanied by two screening
$t_0$-charges for the same reason.
The resulting configuration should thus resemble Fig.~\ref{fig:cg_gp}(a).
The leading contribution to the Green function will then come from the
residual interaction of these partially screened d-charges,
whose charges are ($\pm 1$ times) $\vec{e}_0-\vec{e}_d = (1/\sqrt{g},0)$.
Thus, for large $\tau-\tau^\prime$ we have 
$G_D (\tau-\tau^\prime) \sim
\text{sgn}(\tau^\prime-\tau) |\tau-\tau^\prime|^{-1/g}$.
This leads to $\rho_D (\omega) \sim |\omega|^{1/g-1}$, the usual
tunneling density of states singularity at the end of a LL wire
\cite{bosonization,kane92}.

When $\varepsilon_0$ is small (with respect to $T_K$),
renormalization effects are significant.
In addition to the usual
CG RG Eq.~(\ref{eqn:rg2})
representing the screening of interaction between $t_0$-charges
by pairs of nearby $t_0$-charges (separated by $\xi$) \cite{yuval_anderson},
one can write down similar equations for the flow of the Green function
parameters \cite{si93}.
It is easy to see
that the coefficients of the logarithmic
interaction between any two charges
(either both $t_0$-charges, both d-charges, or a mixed pair)
are renormalized in the same way by the pairs of nearby $t_0$-charges,
i.e.,
\begin{equation}
 \frac{d (\vec{e}_\mu \cdot \vec{e}_\nu)}{d \ln \xi} =
 -2 y^2 \left( \vec{e}_\mu \cdot \vec{e}_0 + \vec{e}_0 \cdot \vec{e}_\nu \right),
\end{equation}
where $\mu,\nu=0,d$.
Thus, the combination
$|\vec{e}_0-\vec{e}_d|^2 =
|\vec{e}_0|^2 + |\vec{e}_d|^{2} - 2 \vec{e}_0 \cdot \vec{e}_d$
is invariant, and retains its initial value of $1/g$.
At the strong coupling fixed point $y$ is large,
so, by Eq.~(\ref{eqn:rg2}),
$|\vec{e}_d^{\thinspace*}|=\sqrt{\alpha_\text{FES}^*}=0$,
where asterisks denote fixed-point values.
Thus, $|\vec{e}_0^{\thinspace*}| = 1/\sqrt{g}$, so that
$G_D (\tau^\prime-\tau) \sim (\tau^\prime-\tau)^{-|\vec{e}_d^{\thinspace*}|^2}
= (\tau^\prime-\tau)^{-1/g}$,
resulting again in $\rho_D (\omega) \sim |\omega|^{1/g-1}$.
Since this behavior holds at both large and small $\varepsilon_0$
values, it should also apply at all intermediate values.

Further support for this result is obtained by
analysis of the particular case $\alpha_\text{FES}=1$
(the Toulouse limit \cite{yuval_anderson}),
where the CG, and all thermodynamic
properties, reduce to those of a noninteracting resonant level
(for which $g=1$, $U_0=0$, and $K=0$).
A nontrivial (i.e., interacting) realization of this condition,
which still permits an exact calculation of the LDoS,
is the case of no coupling to a bath ($K=0$), but with $g=1/4$ and
a corresponding compensating value of the dot-lead interaction.
Then $\vec{e}_d = (-1,0) = - \vec{e}_0$, so
the d-charges have the same magnitudes as the corresponding
$t_0$-charges, but the opposite signs.
Comparison of the corresponding CG expansions thus
shows that the level Green function of the interacting
system is equal to a two-particle Green function of the
noninteracting resonant level
$\langle \hat{T}_\tau \psi(0,\tau) d(\tau)
d^\dagger(\tau^\prime) \psi^\dagger(0,\tau^\prime)  \rangle$,
up to a factor of $(2\pi a)^{-1}\text{sgn}(\tau^\prime-\tau)$.
The operators at $\tau$ and $\tau^\prime$ in this noninteracting two-particle
Green function are similar to the $t_0$ term in the Hamiltonian,
but with $\psi(0)$ replaced by $\psi^\dagger(0)$ to account
for the signs of the d-charges in the interacting system.
After a straightforward evaluation of this two particle Green function
by Wick's theorem, we find for the LDoS the following expression:
\begin{multline}
\rho_D(\omega) = 2 \rho_L a 
\coth \left( \frac{\omega}{2T} \right)
\text{Im} \left\{
\frac{\omega - 2 \varepsilon_0 + 2 i \Gamma_0}{\omega - 2 \varepsilon_0}
\times \right. \\ \left.
\left[
\psi \left( \frac{1}{2} -
\frac{\varepsilon_0 - i \Gamma_0}{2 \pi i T} \right)
- \psi \left( \frac{1}{2} +
\frac{\omega - \varepsilon_0 + i \Gamma_0}{2 \pi i T} \right)
\right]
\right\},
\end{multline}
where $\psi(z)$ is the digamma function \cite{absteg}.
For small $|\omega|$ and $T$ we indeed recover the $[\max(T,|\omega|)]^3$ behavior
appropriate for $g=1/4$, for \emph{all} values of $\varepsilon_0$.

Physically, the result is clear: in the strongly-coupled phase
the level behaves, at low energies, as the last site of the lead,
so its LDoS is similar to the local density
of states near the end of a LL wire \cite{bosonization,kane92},
i.e., $\rho_D (\omega) \sim |\omega|^{1/g-1}$.
Interestingly, not only dot-lead interactions,
but even coupling to the bath does not modify this behavior.
As a result of this, the LDoS
exhibits a power-law behavior with exponent which depends only on
the LL parameter $g$ and not on $\alpha_{\text{FES}}$, i.e.,
on the interactions in the wire but
not on the level-lead and level-bath couplings.
This is in contrast with, e.g., the level occupancy,
which depends on $\alpha_{\text{FES}}$ but not on $g$,
as discussed above. Below we also test these predictions numerically.

\subsection{The weak-coupling phase}

We now turn our attention to the weak-coupling phase.
Here all the $t_0$-charges
are bound in pairs, so the high $|\varepsilon_0|$ results discussed above
(which should also hold in this phase)
actually carry over to low values of $|\varepsilon_0|$. It should however
be remembered that then they compete with the contribution of
the $t_0=0$ term in the CG expansion (the weak coupling fixed point),
the term representing the interaction of the unscreened d-charges,
which gives the LDoS a contribution of the form
$\rho_\text{pair} (\omega) \sim |\omega|^{\alpha_\text{FES}^d-1}$,
with $\alpha_\text{FES}^d \equiv |\vec{e}_d|^2$.
This is simply the LDoS of a tunnel-decoupled level,
broadened from a delta peak to a power-law by the Anderson orthogonalities
in the wire \cite{noziers69} and in the Ohmic bath \cite{spin_boson,spin_boson2}.
Note that since the level is effectively decoupled,
the usual $|\omega|^{1/g-1}$ behavior at the end of a LL wire
need not apply anymore.

One could actually proceed to study higher order terms
in the weak-coupling regime
(and similarly, for large $|\varepsilon_0|$ in the strong-coupling phase).
For small $t_0$ the leading correction is
dressing of the above-mentioned charge configurations by
a series of pairs of close-by $t_0$-charges
(close-by since $|\varepsilon_0|$ and/or $\alpha_\text{FES}$ are large),
as depicted in Fig.~\ref{fig:cg_gp}.
One then has to sum over all the terms similar to
Fig.~\ref{fig:cg_gp}(a) for large $|\varepsilon_0|$
(i.e., for $|\omega| \ll |\varepsilon_0|$ in the weak-coupling phase)
or all the terms
similar to Fig.~\ref{fig:cg_gp}(b) for small $|\varepsilon_0|$
(i.e., for $|\omega| \gg |\varepsilon_0|$).
Since each pair has a very small dipole moment
(due to the proximity of the charges),
inter-pair interactions are negligible in a first
approximation.
This is actually an imaginary time variant of the
noninteracting blip approximation (NIBA) \cite{spin_boson,spin_boson2}.
The following argument can spare us the need of explicit calculations.
For $U_0=0$ and $K=0$, the d-charges are noninteracting, $\vec{e}_d=(0,0)$.
The sum over all the terms with pairs of nearby $t_0$ charges
would be the same (in the current approximation)
as the Green function of a \emph{noninteracting}
system consisting of a level tunnel-coupled to a bath of
\emph{noninteracting} fermions with power-law local density of
states $\rho_L (\omega) \sim |\omega|^{\alpha_\text{FES}-1}$
(with some appropriate high-energy cutoff and normalization).
For the latter system the Green function can be easily evaluated
to give
$G_D^{0} (i\omega) = [i\omega - \varepsilon_0-\Sigma_D^{0} (i\omega)]^{-1}$,
where the dot self energy is \cite{mahan}:
\begin{equation}
 \Sigma_D^{0}(i\omega) = \left( t_0 \right)^2
 \int_{-\infty}^\infty \frac{\rho_L(\Omega)}{i\omega - \Omega} d \Omega,
\end{equation}
so that $\Sigma_D^{0}(i\omega) \sim \omega^{\alpha_\text{FES}-1}$.
For small $|\varepsilon_0|$ we indeed see that $\Sigma_D^0(\omega)$
is subdominant with respect to the noninteracting contribution
only if $\alpha_\text{FES}>2$, which is exactly the condition for the
weak-coupling phase for small $t_0$.
Then, exactly at $\varepsilon_0=0$, a delta-function term
appears at $\omega=0$ in the expression for the LDoS (similarly to the
situation at $t_0=0$), whose coefficient is determined by
the requirement that the integral of the entire expression
for the LDoS corresponding to $G_D^{0}(i\omega)$ is unity \cite{furusaki02}.

Before discussing nonzero $U_0$ and $K$, it should be remarked that
this NIBA-like approximation exactly reproduces the perturbative
(in the tunneling $t_0$) approach employed in Ref.~\onlinecite{wachter07},
and would lead to similar predictions for the behavior of the level
population.
Both approximations are justified only in the weak-coupling phase
(or when $|\varepsilon_0|$ is large), but not in the strong-coupling
phase, where $t_0$ grows under RG flow and thus cannot be treated
perturbatively, similarly to the exchange $J_{xy}$
in the equivalent Kondo problem \cite{hewson}.
In the strong-coupling regime perturbative results predict correctly
the dependence of the Kondo temperature on $t_0$
(again, just like in the Kondo model \cite{hewson})
and the qualitative behavior of
the LDoS for $1/2<g<1$ and $\varepsilon_0 \ne 0$ (at $U_0=0$ and $K=0$),
but deviate from our previous conclusions in many other respects.
For example, as we discuss below, at nonzero $U_0$ and $K$ our NIBA
calculations indicate that the exponent in the power-law behavior
of the LDoS at low energy may depend on these interactions too, in
contrast with the situation in the strong-coupling phase,
where the corresponding
exponent depends only on the LL parameter $g$, as shown above.
Moreover, even for vanishing dot-lead and dot-bath interactions
(the case treated in Ref.~\onlinecite{wachter07})
the NIBA/perturbative expression given above does not agree with our
previous analysis of the strong-coupling phase:
for (a) $1/2<g<1$ and $\varepsilon_0=0$ or (b) $g>1$ and any
$\varepsilon_0$, NIBA would suggest that the LDoS varies
as $\sim |\omega|^{1-1/g}$, i.e., with the \emph{opposite} exponent
to the one appearing in the density of states at the end of a
LL wire. Perturbative results would thus imply that the
LDoS may be \emph{enhanced} for $g<1$ or \emph{suppressed} for
$g>1$, which is clearly at odds with both our previous results
and the behavior of density of states at the lead edge.
Moreover, integrating the perturbative LDoS leads to
the prediction that the level population may have a power-law
dependence on $\varepsilon_0$ at the strong-coupling phase
(for $1/2<g<2/3$)\cite{wachter07}, which is in contrast with
the analytical behavior expected from the exact mapping of
our model onto the Kondo problem, as discussed above.
To summarize, NIBA/perturbative (in $t_0$) expressions do not
hold in general in the strong-coupling phase.
It may be noted that the
numerical data of Ref.~\onlinecite{wachter07} does not
cover these regimes of the strong coupling phase for which
perturbative calculations disagree with our previous analysis.

Returning to the discussion of the NIBA approximation
for the weak-coupling phase,
we will now treat the more general case, i.e.,
nonzero $U_0$ and $K$
(which was not addressed in Ref.~\onlinecite{wachter07}).
Again, screened d-charges terms [Fig.~\ref{fig:cg_gp}(a)]
are dominant for $|\omega| \ll |\varepsilon_0|$, whereas
unscreened d-charges terms [Fig.~\ref{fig:cg_gp}(b)]
are dominant for $|\omega| \gg |\varepsilon_0|$.\cite{fn:dominant_term}
Let us start from the latter case.
Now that there is interaction between d-charges,
the CG expression for the Green function $G_D(\tau-\tau^\prime)$
contains an additional
a factor of the form $\sim |\tau-\tau^\prime|^{-\alpha_\text{FES}^d}$
(interaction of d-charges with the pairs of close-by
$t_0$-charges is negligible).
Turning to the frequency domain, the LDoS
of the unscreened d-charges contribution [Fig.~\ref{fig:cg_gp}(b)]
will thus be the convolution of the
LDoS $\rho_D^0 (\omega)$ associated with $G_D^{0}(i\omega)$
from the previous paragraph, with a function
$\rho_{\text{pair}} (\omega) \sim |\omega|^{\alpha_\text{FES}^d-1}$,
which is simply the LDoS of a decoupled ($t_0=0$) level
as discussed above, i.e.,
\begin{equation} \label{eqn:niba}
 \rho_D (\omega) = 2 \text{sgn}(\omega) \int_0^\omega
 \rho_D^0 (\omega - \Omega) \rho_{\text{pair}} (\Omega) d\Omega,
\end{equation}
at $T=0$.
Thus, the $|\omega|^{\alpha_\text{FES}^d-1}$ behavior
of the decoupled level will survive for
vanishing $|\varepsilon_0|$, due to the delta peak in $G_D^{0}(i\omega)$.
This behavior actually applies to the entire
$|\omega| \gg |\varepsilon_0|$ region,
which is exactly where the contribution
of terms similar to Fig.~\ref{fig:cg_gp}(b) is important.
Higher order corrections 
will give extra powers of
$|\omega|^{\alpha_\text{FES}-2}$, which are subleading
since $\alpha_\text{FES}>2$.

\begin{figure*}
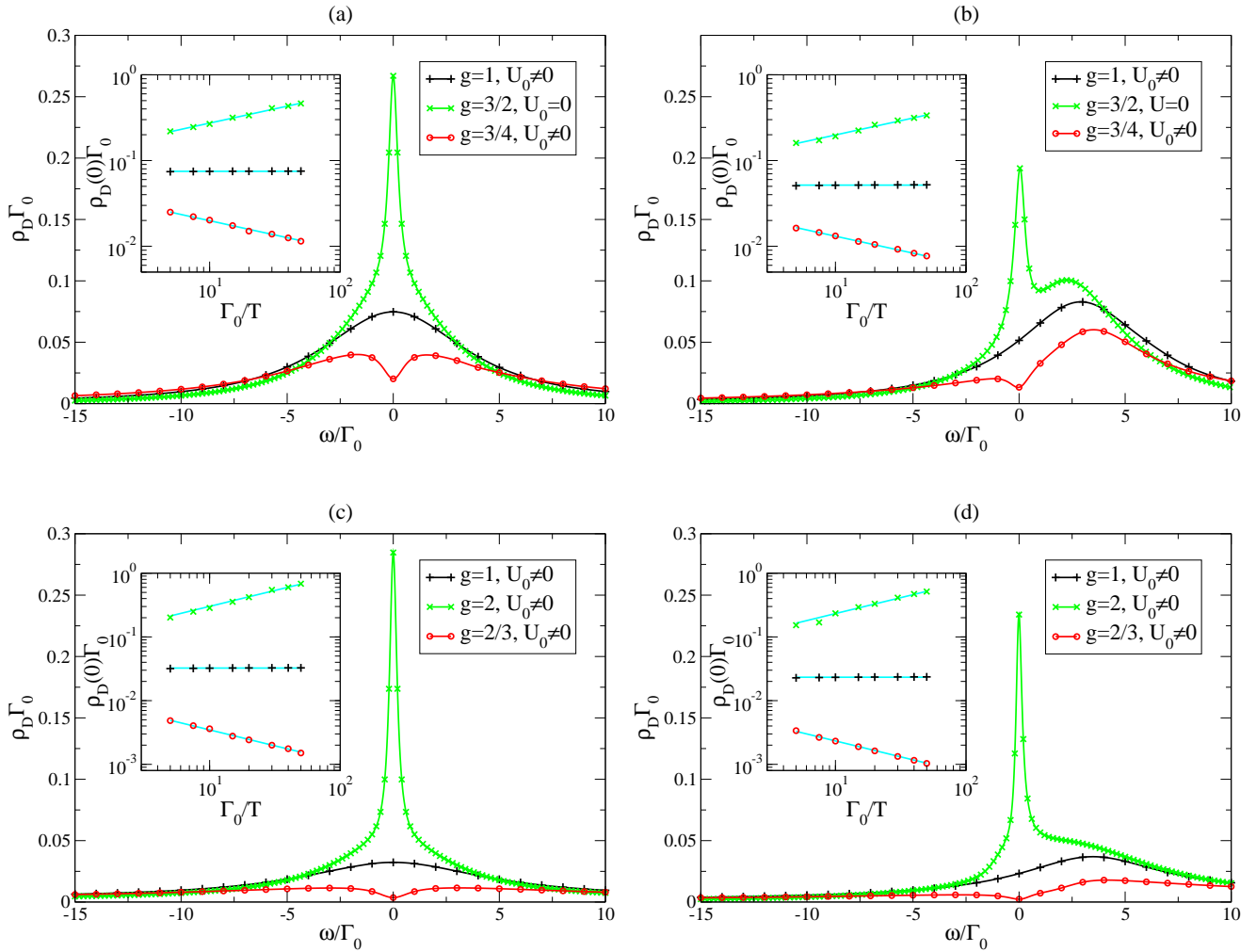

\centering
\begin{tabular}{p{8.6cm}p{8.6cm}}
\includegraphics[width=8.5cm,height=!]{ldos_e0K0} &
\includegraphics[width=8.5cm,height=!]{ldos_e2gK0} \\[.6cm]
\includegraphics[width=8.5cm,height=!]{ldos_e0KK} &
\includegraphics[width=8.5cm,height=!]{ldos_e2gKK}
\end{tabular}
\caption{\label{fig:ldos}
(Color online)
MC results for the LDoS as a function of frequency (measured with respect to the
Fermi energy) for three different models, all with $\alpha_{\text{FES}}=2/3$
and $T/\Gamma_0=0.1$.
The three curves in each panel correspond to three different values of $g$:
$g=1$ (black pluses), $g>1$ (green X's), and $g<1$ (red circles),
as indicated in the legends.
The insets present the the LDoS at the Fermi energy as a function of the temperature
by symbols (with the same code as before),
together with a cyan line showing best-fit to the expected power-law behavior
$\rho_D(0) \sim T^{1/g-1}$ (only the prefactor is fitted).
The different panels correspond to different values of $\varepsilon_0$ and $K$:
(a) $\varepsilon_0 = 0$ and $K=0$;
(b) $\varepsilon_0 = 2\Gamma_0$ and $K=0$;
(c) $\varepsilon_0 = 0$ and $K=1/6$;
(d) $\varepsilon_0 = 2\Gamma_0$ and $K=1/6$.
In each case the value of $U_0$ was chosen
according to the requirement $\alpha_{\text{FES}}=2/3$.
}
\end{figure*}

Similar considerations apply to the screened d-charges contribution
[Fig.~\ref{fig:cg_gp}(a)]
when $|\omega| \ll |\varepsilon_0|$.
For $U_0 \ne 0$ and/or $K \ne 0$, two corrections are due.
The first correction takes into account the factors coming from the
interaction between each d-charge and the neighboring $t_0$-charge,
which are power-laws in the time domain.
Since the d-$t_0$ charges form tightly-bound pairs,
we can take the limits of integration over their separation
to infinity. They thus yield factors of
\begin{equation}
 \int_{0}^\infty
 \left( \frac{\xi}{\tilde{\tau}} \right)^{\vec{e}_0 \cdot \vec{e}_d}
 e^{-|\varepsilon_0| {\tilde{\tau}}} \frac{d{\tilde{\tau}}}{\xi} =
 \frac{\Gamma (1 - \vec{e}_0 \cdot \vec{e}_d)}
 { \left( |\varepsilon_0| \xi \right) ^ {1 - \vec{e}_0 \cdot \vec{e}_d} },
\end{equation}
where $\Gamma(z)$ is the gamma function \cite{absteg}, and
$\vec{e}_0 \cdot \vec{e}_d = (\alpha^d_\text{FES} + \alpha_\text{FES} - 1/g)/2$.
The correction is then the ratio between this expression
and its value at $\vec{e}_d = 0$.
Apart from this constant factor,
one must compensate for the fact
that the inter-pair interaction of these two d-$t_0$ pairs
gives the Green function a factor which varies as
$|\tau-\tau^\prime|^{-1/g}$
(since $|\vec{e}_d - \vec{e}_0|^2 = 1/g$), instead of
the $|\tau-\tau^\prime|^{-\alpha_\text{FES}}$
dependence used in the calculation of $G_D^0$.
Hence, $\rho^0_D(\omega)$
should be convoluted here with
$\sim |\omega|^{1/g-\alpha_\text{FES}-1}$.
For $|\omega| \ll |\varepsilon_0|$ the LDoS
will thus retain the $|\omega|^{1/g-1}$
behavior of the strong-coupling phase.

To conclude the discussion of this NIBA-type approximation,
a general expression for the LDoS which interpolates
between large and small $|\omega/\varepsilon_0|$ limits \cite{fn:dominant_term}
is of the form of Eq.~(\ref{eqn:niba}),
but with $\rho_{\text{pair}} (\Omega)$
replaced by
\begin{equation}
 \rho^\prime_{\text{pair}} (\Omega) \sim
 |\Omega|^{\alpha_\text{FES}^d-1}
 \left\vert \frac{\varepsilon_0}{\Omega} - 1
 \right\vert^{\alpha_\text{FES}+\alpha_\text{FES}^d-1/g}.
\end{equation}
All the low-energy results of this section
are summarized in Table~\ref{tbl:rho_d}.

\section{Numerical calculations} \label{sec:numerics}
In this section we present the results of numerical calculations,
verifying the conclusions of our previous analysis, i.e.,
that the LDoS at low energies features a power-law behavior with
the power determined by LL physics only (in the strong coupling phase),
although, as we have shown before \cite{goldstein09}, 
its integral (the level population) is universal, and
cannot be used to extract LL parameters.

To calculate the LDoS we used classical Monte-Carlo (MC) simulations
on the CG expansion of dot Green function \cite{schotte71}.
The MC update procedure used is similar to the one employed recently
for the closely-related continuous time quantum MC algorithm \cite{werner06}.
After obtaining the imaginary-time Green function
it was Fourier-transformed to Matsubara frequencies,
followed by analytic continuation to real
frequencies using the Pad\'{e} approximant technique \cite{pade}.
This yields the retarded Green function, whose imaginary part
is proportional to the LDoS \cite{mahan}.
Below we present data in the non-perturbative strong-coupling region,
which confirms the results of our previous analysis.
Actually, in the weak-coupling phase
(which is accessible analytically through the NIBA-like approximation)
MC simulations are not efficient, since
there CG charges are rare and averaging very slow.
In this sense, our analysis and numerical calculations are complementary.

The results presented in the different panels Fig.~\ref{fig:ldos}.\
The values of $g$, $U_0$ and $K$ are varied, in a way which keeps
$\alpha_{\text{FES}}$ constant at a value of $2/3$.
Hence, the occupancies as functions of $\varepsilon_0$ are
the same, as we also verify below (actually, the CG
representation would predict exactly identical occupations).
The LDoS curves are, however, markedly different:
depending on whether $g>1$, $g<1$, or $g=1$, they
have a maximum, a minimum, or no special feature near the
Fermi energy, respectively.
In the inset we demonstrate that in all cases the LDoS at the Fermi
energy exhibits power-law dependence on temperature,
$\rho_D(0) \sim T^{1/g-1}$, as found in the previous section
(cf.\ Table~\ref{tbl:rho_d}).

\begin{figure}
\includegraphics[width=8.5cm,height=!]{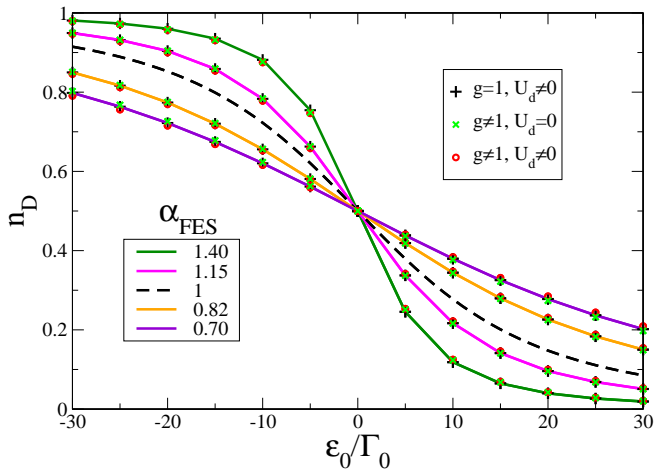}
\caption{\label{fig:nlevel}
(Color online)
DMRG results for the average level occupancy $n_D$ as a function of
its energy, $\varepsilon_0$,
for a discrete model of the lead [Eqs.~(\ref{eqn:h_tb1})--(\ref{eqn:h_tb2})].
The curves on which the symbols reside (which serve as guides to the eye)
correspond to the various $\alpha_{\text{FES}}$ values, where
the larger $\alpha_{\text{FES}}$ the narrower the curve and vice versa.
The widest curve has similar parameters
to those used in the MC simulations presented in Fig.~\ref{fig:ldos}.
On each curve there are three different choices of $U$ and $U_d$
(all giving the same $\alpha_{\text{FES}}$ value),
denoted by the three different symbol types.
See the text for further details.
}
\end{figure}

For the sake of completeness, we will
repeat here some of our previous data on the level occupancy \cite{goldstein09}.
Since the MC simulation is based on the CG, to have an independent
check of the universality of the level population
we employed
the density matrix renormalization group
(DMRG) \cite{dmrg} algorithm, using block-sizes of up to $256$.
DMRG is also better suited to ground state calculations,
and thus complements the necessarily finite-temperature MC in this
respect.
The model used is a
half-filled tight binding chain with nearest-neighbor interactions.
It is described by the Hamiltonian:
\begin{equation}
\label{eqn:h_tb1}
 H_{L} = 
    \sum_{i=1}^{N-1} \left[
    t c_{i}^{\dagger} c_{i+1} +  \text{H.c.}
   +
   U \left( c_{i}^{\dagger} c_{i} - \textstyle\frac{1}{2} \right)
     \left( c_{i+1}^{\dagger} c_{i+1} - \textstyle\frac{1}{2} \right)
   \right] 
\end{equation}
where $c^{\dagger}_{i}$ ($c_i$) is the electronic creation (annihilation)
operator at the $i$th site of the wire ($i=1 \ldots N$, with $L = N a$),
and $t$ and $U$ are, respectively, nearest-neighbor hopping amplitude
and interaction strength.
The low energy physics of this model is known to be governed by LL
theory for not too large interactions (i.e., $|U|<2t$),\cite{bosonization}
with $g=\pi/[2\cos^{-1}(-U/2t)]$
and $v/(2 a t)=\pi \sqrt{1-(U/2t)^2}/[2 \cos^{-1}(U/2t)]$.
The dot is still governed by the same $H_D=\varepsilon_0 d^\dagger d$,
and is coupled to the lead through:
\begin{equation} \label{eqn:h_tb2}
 H_{DL} = 
   t_d c_1^{\dagger} d + \text{H.c.}
   +
   U_d
     \left( d^{\dagger} d - \textstyle\frac{1}{2} \right)
     \left( c_{1}^{\dagger} c_{1} - \textstyle\frac{1}{2} \right)
\end{equation}
$t_d$, $U_d$ are related to the corresponding parameters of the continuum
version Eq.~(\ref{eqn:hdl}) by
$t_0 = t_d \sqrt{a}$, and
$U_0 = U_d a$,
$a$ being the lattice spacing.
We have previously shown that boundary conformal field theory arguments
and the Bethe ansatz solution yield that here
$\delta_{\text{eff}}=\tan^{-1} ( { U_d } /
{ \sqrt{4t^2-U^2} } )$. \cite{goldstein09}

The level population is plotted in Fig.~\ref{fig:nlevel} as
a function of $\varepsilon_0$.
Different curves correspond to different values of $\alpha_{\text{FES}}$,
as indicated in the legend.
On each such curve there are three types of symbols, denoting DMRG
data on three different models:
(i) $g=1$ (i.e., $U=0$) but nonzero $U_d$;
(ii) $g \ne 1$ (nonzero $U$) but $U_d = 0$;
(iii) both $U$ and $U_d$ are nonzero.
All the models are without coupling to the bath ($K=0$).
The values of $U$ and $U_d$ in each model were chosen so as to give the
same value of $\alpha_{\text{FES}}$ for each curve.
For model (iii) we used $U=\pm 0.5t$, with sign opposite to that of model (ii).
In all cases we chose $t_d$ to get $\Gamma_0=10^{-4}t$
and used $N=100v/(a t)$ sites.
The results clearly show that the occupancy is indeed universal,
depending only on $\alpha_{\text{FES}}$, and not on the
strengths or signs of the interactions ($U$ and $U_d$).
It should be noted that the widest curve has similar parameters
to those used in the MC simulations (Fig.~\ref{fig:ldos}).

\section{Conclusions} \label{sec:conclude}

To summarize, we have studied, both analytically and numerically,
the LDoS of a level coupled to a LL and to an Ohmic bath
over the entire parameter space.
We have found that in general it exhibits a power-law dependence
at low energies.
In large parts of the phase space this is just
the power-law behavior of the tunneling density of states
at the end of a LL wire.
Thus, a measurement of the LDoS there can be used
to extract the value of the LL interaction parameter $g$.
In other regions it is also
affected by level-lead and level-bath interactions.
In any case the LDoS is explicitly sensitive to the value
of the LL parameter $g$, although the LDoS
determines the level population,
which was found before to be universal \cite{goldstein09}, 
and thus not to feature any LL-specific power-law.

\begin{acknowledgments}
We would like to thank Y. Weiss for his invaluable
help with the DMRG calculations,
A. Schiller for many useful suggestions,
and Y. Gefen for discussions.
M.G. is supported by the Adams Foundation of the Israel Academy
of Sciences and Humanities.
Financial aid from the Israel Science Foundation (Grant 569/07) is
gratefully acknowledged.
\end{acknowledgments}


\begin{thebibliography}{99}

\bibitem{bosonization}
A.O. Gogolin, A.A. Nersesyan, and A.M. Tsvelik,
\textit{Bosonization and Strongly Correlated Systems},
(Cambridge University Press, Cambridge, 1998);
T. Giamarchi, \textit{Quantum Physics in One Dimension},
(Oxford University Press, Oxford, 2003).

\bibitem{chang03}
For a review see: A.M. Chang, \rmp \textbf{75}, 1449 (2003),
and references cited therein.

\bibitem{koenig08} For a review see:
M. K\"{o}nig, H. Buhmann, L.W. Molenkamp, T. Hughes, C.-X. Liu, X.-L. Qi, and S.-C. Zhang,
J.\ Phys.\ Soc.\ Jpn.\ \textbf{77}, 031007 (2008),
and references cited therein.

\bibitem{kane92}
C.L. Kane and M.P.A. Fisher, \prl \textbf{68}, 1220 (1992);
\prb \textbf{46}, R7268 (1992); \textbf{46}, 15233 (1992).

\bibitem{furusaki93} A. Furusaki and N. Nagaosa, \prb \textbf{47}, 3827 (1993).

\bibitem{furusaki98}
A. Furusaki, \prb \textbf{57}, 7141 (1998).

\bibitem{auslaender00}
O.M. Auslaender, A. Yacoby, R. de Picciotto, K.W. Baldwin, L.N. Pfeiffer, and K.W. West,
\prl \textbf{84}, 1764 (2000).

\bibitem{postma01}
H.W.Ch. Postma, T. Teepen, Z. Yao, M. Grifoni, and C. Dekker,
Science \textbf{293}, 76 (2001).

\bibitem{komnik03}
A. Komnik and A.O. Gogolin, \prl \textbf{90}, 246403 (2003);
\prb \textbf{68}, 235323 (2003).

\bibitem{nazarov03}
Yu.V. Nazarov and L.I. Glazman, \prl \textbf{91},
126804 (2003).

\bibitem{polyakov03}
D.G. Polyakov and I.V. Gornyi, \prb \textbf{68},
035421 (2003).

\bibitem{lerner08}
I.V. Lerner, V.I. Yudson, and I.V. Yurkevich,
\prl \textbf{100}, 256805 (2008).

\bibitem{goldstein10b}
M. Goldstein and R. Berkovits,
\prl \textbf{104}, 106403 (2010).

\bibitem{furusaki02}
A. Furusaki and K.A. Matveev,
\prl \textbf{88}, 226404 (2002).

\bibitem{sade05}
M. Sade, Y. Weiss, M. Goldstein, and R. Berkovits,
\prb \textbf{71}, 153301 (2005);


\bibitem{lehur05}
K. Le Hur and M.-R. Li, \prb \textbf{72}, 073305 (2005).

\bibitem{weiss06}
Y. Weiss, M. Sade, M. Goldstein, and R. Berkovits,
Phys. Stat. Sol. (b) \textbf{243}, 399 (2006);
Y. Weiss, M. Goldstein, and R. Berkovits,
J.\ Phys.: Condens.\ Matter  \textbf{19}, 086215 (2007);

\bibitem{weiss07}
Y. Weiss, M. Goldstein, and R. Berkovits,
\prb \textbf{75}, 064209 (2007); 
\textbf{76}, 024204 (2007); 
\textbf{77}, 205128 (2008).

\bibitem{wachter07}
P. W\"{a}chter, V. Meden, and K. Sch\"{o}nhammer,
\prb \textbf{76}, 125316 (2007).

\bibitem{bishara08}
G.A. Fiete, W. Bishara, C. Nayak,
\prl \textbf{101}, 176801 (2008);
\prb \textbf{82}, 035301 (2010).

\bibitem{goldstein09}
M. Goldstein, Y. Weiss, and R. Berkovits,
Europhys.\ Lett.\ \textbf{86}, 67012 (2009);
Proceedings of FQMT `08,
Physica E \textbf{42}, 610 (2010).

\bibitem{elste10}
F. Elste, D.R. Reichman, and A.J. Millis, \prb \textbf{81}, 205413 (2010).

\bibitem{fn:fabrizio95}
A non-chiral LL with a boundary can mapped onto a chiral LL by unfolding the
decoupled (Bogolubov-transformed) right- and left-movers
[M. Fabrizio and A.O. Gogolin, \prb \textbf{51}, 17827 (1995)].

\bibitem{spin_boson}
A.J. Leggett, 
S. Chakravarty, A.T. Dorsey, M.P.A. Fisher, A. Garg, and W. Zwerger,
\rmp \textbf{59}, 1 (1987).

\bibitem{spin_boson2}
U. Weiss, \textit{Quantum Dissipative Systems} (World Scientific, Singapore, 1999).

\bibitem{kamenev_gefen}
A. Kamenev and Y. Gefen, \prb \textbf{54}, 5428 (1996);
arXiv:cond-mat/9708109. 

\bibitem{yuval_anderson}
P.W. Anderson and G. Yuval, \prl \textbf{23}, 89 (1969);
G. Yuval and P.W. Anderson, \prb \textbf{1}, 1522 (1970);
P.W. Anderson, G. Yuval, and D.R. Hamann,
\textit{ibid.} \textbf{1}, 4464 (1970).

\bibitem{schotte71}
K.D. Schotte and U. Schotte,
\prb \textbf{4}, 2228 (1971).

\bibitem{wiegmann78}
P.B. Wiegmann and A.M. Finkelstein,
Zh.\ Eksp.\ Teor.\ Fiz.\ \textbf{75}, 204 (1978)
[Sov.\ Phys.\ JETP \textbf{48}, 102 (1978)].

\bibitem{si93}
Q. Si, and G. Kotliar, \prb \textbf{48}, 13881 (1993).

\bibitem{fabrizio95}
M. Fabrizio, A.O. Gogolin, and P. Nozi\`{e}res,
\prb \textbf{51}, 16088 (1995).

\bibitem{noziers69}
P. Nozi\`{e}res and C.T. De Dominicis, Phys. Rev. \textbf{178}, 1097 (1969).

\bibitem{mahan}
G.D. Mahan, \textit{Many-Particle Physics}
(Kluwer, New York, 2000). 

\bibitem{hewson}
A.C. Hewson, \textit{The Kondo Problem to Heavy Fermions}
(Cambridge University Press, Cambridge, 1993).

\bibitem{fn:dominant_term}
When $1/g < \alpha^d_\text{FES}$ (i.e., in the presence
of strong dot-lead and/or dot-bath interactions),
interaction between d-charges is large enough to make
the screened d-charges contribution [Fig.~\ref{fig:cg_gp}(a)]
dominant even for small $|\varepsilon_0|$.
Using similar methods one can show that then
the LDoS will vary as $|\omega|^{1/g-1}$,
just like for $|\omega| \ll |\varepsilon_0|$.

\bibitem{absteg}
M. Abramowitz and I.A. Stegun,
\textit{Handbook of Mathematical Functions} (Dover, New York, 1964).

\bibitem{werner06}
P. Werner, A. Comanac, L. de' Medici, M. Troyer, and A.J. Millis,
\prl \textbf{97}, 076405 (2006);
P. Werner and A.J. Millis, \prb \textbf{74}, 155107 (2006).

\bibitem{pade}
H.J. Vidberg and J.W. Serene, J.\ Low Temp.\ Phys.\ \textbf{29}, 179 (1977).

\bibitem{dmrg}
U. Schollw\"{o}ck, \rmp \textbf{77}, 259 (2005);
K.A. Hallberg, Adv.\ Phys.\ \textbf{55}, 477 (2006).


\end{thebibliography}
\end{document}